\documentclass[runningheads,a4paper]{llncs}  

\usepackage{geometry}                		% See geometry.pdf to learn the layout options. There are lots.
\usepackage[pdftex]{graphicx}
\usepackage[justification=centering]{caption}
\usepackage{comment}
\usepackage{url}
\usepackage{amssymb,amsmath}
\usepackage{array}
\newcolumntype{L}{>{\centering\arraybackslash}m{3cm}}
\urldef{\mailsa}\path|{sadara@cisco.com,murali@iiitb.ac.in}|

\begin{document}

% paper title
\title{Privacy Preserving Architectures for Collaborative Intrusion Detection \\ {\small Position Paper} }

\author{Sashank Dara\inst{1} \and Dr.V.N Muralidhara\inst{2}}

\institute{Cisco Systems Inc \& IIIT-Bangalore \\
Bangalore, 560037, India\\
\email{sadara@cisco.com} \\
\and
IIIT-Bangalore\\
Bangalore, 560100, India\\
\email{murali@iiitb.ac.in}}

% author names and affiliations
% use a multiple column layout for up to three different
% affiliations
\begin{comment} %IEEE formats
\author{
\authorblockN{Sashank Dara}
\authorblockA{Cisco Systems Inc\\
IIIT-Bangalore \\
Bangalore, 560037, India\\
sadara@cisco.com}  \and
\authorblockN{Dr.V.N Muralidhara}
\authorblockA{IIIT-Bangalore\\
Bangalore, 560100, India\\
murali@iiitb.ac.in}
}
\end{comment}
% avoiding spaces at the end of the author lines is not a problem with
% conference papers because we don't use \thanks or \IEEEmembership
% for over three affiliations, or if they all won't fit within the width
% of the page, use this alternative format:
% make the title area
\maketitle

\begin{abstract}
Collaboration among multiple organizations is imperative for contemporary intrusion detection. As modern threats become well sophisticated it is difficult for organizations to defend with threat context local to their networks alone. Availability of global \emph{threat intelligence} is must for organizations to defend against modern advanced persistent threats (APTs). In order to benefit from such global context of attacks, privacy concerns continue to be of major hindrance. In this position paper we identify real world privacy problems as precise use cases, relevant cryptographic technologies and discuss privacy preserving architectures for collaborative intrusion detection.

\end{abstract}

\section{Introduction}
Contemporary intrusion detection is very daunting and challenging task due to two major factors. Firstly  network infrastructures are becoming quite complex due to rapid technology advances\footnote{ Due to areas like Cloud, BOYD, SDN/NFV, IOT etc.}. Secondly modern cyber security attacks are very advanced, persistent and devastating. Threat actors are well funded, organized and such attacks happen at a global scale. Variety of information is needed to make modern intrusion detection possible and collaboration among different organizations is very imperative. 

\subsection{Telemetry}

Identification of attacks would need \emph{Telemetry}. This includes browsing history, network flows, file attributes, packet captures, firewall logs, application logs, system level behavior attributes etc. Verbose telemetry would facilitate event correlation, machine learning, security analytics and other techniques for identifying intrusions. %Such efforts help security researchers in learning attack patterns, malware propagation,

\subsection{Threat Intelligence}
Availability of telemetry will help in tracing specific attack artifacts like known bad domains, URLs serving malware, blacklisted  IP addresses, malicious file signatures etc. Such precise attack artifacts are called \emph{Indicators of Compromise(IOCs)} and are considered crucial for intrusion detection \cite{ioc21ct},\cite{iocsans}. 

  \emph{IOC}s together with other security data like IP reputations, user reputations, vulnerabilities, signatures etc are broadly termed as \emph{Threat Intelligence(TI)}\cite{ti}. It is provided by various entities like commercial vendors, open source communities, government agencies, consortia etc.\cite{tx}
  
\begin{comment}
Simplistic definition is 
\begin{quote}
\emph{Threat Intelligence(TI)} is security data that provides the ability to
prepare to detect, prevent, or investigate emerging attacks
before your organization is attacked\cite{ti}.
\end{quote}  
\end{comment}

\subsection{Privacy Challenges}
Organizations would need to collaborate with their peers, threat intelligence providers and some times even other victims in order to defend against modern threats that happen at global scale\cite{gilbert2014scalable},\cite{zetter2010google}. For such collaborative efforts to be successful it is essential for organizations to share their local telemetry\footnote{ More concrete use cases are provided in detail later}. Sharing telemetry would severely compromise the privacy of individuals and organizations. Often telemetry would contain 

\begin{enumerate}
\item Sensitive \emph{Personally Identifiable Information (PII)} like email-ids, user names, addresses, SSNs, credit card numbers, device identifiers like internal IP addresses etc.
\item \emph{Protected Health Information (PHI)} like health related websites visited, health records, test reports etc.
\item Classified and confidential information like sensitive documents, trade secrets etc.  
\end{enumerate}
Many organizations are constrained to participate in such collaborative efforts due to privacy compliance, regional laws, risk of information leakage etc. and suffer from lacking global context of threats \cite{rsareport}.

Contemporary privacy preserving methods are predominantly based on  anonymization techniques,  they have been well studied and determined to be prone to inferential attacks and other short comings \cite{pang2006devil},\cite{coull2008taming},\cite{king2009taxonomy},\cite{li2009tradeoff},\cite{brekne2005circumventing}.

\subsection{ Key Contributions}
Although much of contemporary body of knowledge is about enhancing modern intrusion detection very little has been published on possible directions for privacy preservation. We believe leveraging the recent advances in cryptographic technologies would aide to better privacy preserving intrusion detection. In this position paper we identify real world problems as use cases, identify possible privacy technologies and architectures, we also discuss their threat models and challenges. We hope our work will motivate practical advances in this emerging field.
\section{Prior Art}
The immense benefit of collaborative distributed intrusion detection was first identified in \cite{locasto2004collaborative}. Subsequent body of work witnessed major advancements in this area \cite{zhou2010survey}.  As the modern attacks became more sophisticated, advanced major research has been done in the areas of big data analytics, anomaly detection, machine learning etc applied for intrusion detection and emerged as very fast emerging field. It is very challenging to summarize these advances but good reference could be found here \cite{suthaharan2014big}. A comprehensive work on Collaborative Intrusion Detection is devoted to advantages of network of multiple intrusion detection systems in order to identify global threats\cite{fung2013intrusion}. Their work focuses very little on privacy issues. Also as we shall discuss in later sections there are many other forms of collaboration that are possible.

Application for \emph{Secure Multiparty Computation} techniques for collaborative intrusion detection were first identified as open problem in \cite{du2001secure}. Later much of the research progressed in privacy preserving data mining techniques\cite{fung2010privacy}. But major break through work in this area of privacy preserving collaborative security monitoring appeared in \cite{burkhart2010sepia}, \cite{burkhart2011privacy}. Subsequently work appeared in privacy preserving network outage monitoring \cite{djatmiko2013collaborative}, anomaly detection \cite{zhang2012privacy}, traffic statistics \cite{brown2013haze}, deep packet inspection\cite{sherryblindbox}, MiTB attacks \cite{qusa2013secure}, predictive blacklists\cite{melis2015building}\cite{freudiger2015controlled}, data leak detection\cite{shu2013data}, information aggregation \cite{kreitz2012practical}, event correlation\cite{qusa2012privacy}. 

There is also exciting advances happening in cryptographic techniques collectively being called as \emph{Computing on Encrypted Data(COED)}. It includes techniques like \emph{Fully Homomorphic Encryption}, \emph{Secure Multiparty Computation}, \emph{Searchable Encryption} etc. A draft version describing these advances could be found here\cite{coed}. We identify relevant techniques needed for privacy preserving intrusion detection in next section.
\section{Preliminaries}\label{section:pri}
In this section we give informal introduction to different privacy preserving techniques in \emph{COED} \footnote{Formal definitions of the same are out of scope for this paper}. Some of these techniques are becoming mature while some are still in nascent stages.
\subsection{Private Information Retrieval}
\emph{Private Information Retrieval (PIR)} techniques are an interesting cryptographic primitive. Informally in a typical \emph{PIR} protocol, a \emph{client} queries a \emph{database server} privately such that the \emph{server} returns the requested records without knowing what they are. As a trivial solution the server could return all the records and let the client chose desired ones locally but this has huge communication complexity. It is first shown in \cite{chor1998private} that non-trivial  \emph{PIR} is impossible when only single database server exists in an information theoretic setting.

\emph{Information Theoretic (IT-PIR)} protocols does not rely on any limitation of computing resources available with an adversary.  These protocols are based on multiple database servers that not all of the servers are colluding with each other \cite{beimel2000reducing}, \cite{goldberg2007improving}, \cite{henry2011practical},\cite{olumofin2012revisiting}. \emph{Computational PIR (CPIR)} are protocols rely on the computing resources available with an adversary for proving security guarantees \cite{chor1997computationally}, \cite{kushilevitz1997replication}, \cite{chang2004single}, \cite{aguilar2014xpire}. %Mention about private lookup \cite{cybernetica2015private}

\subsection{Secure Multiparty Computation}
\emph{Secure Multiparty Computation(SMC)} techniques enable multiple parties to compute some public function without revealing their individual inputs. Secure computations are carried out in a distributed manner on \emph{secret shares} of the input data, the results are constructed back without ever revealing the inputs to participants. Recent advances in this technology are promising with open source libraries and commercial offerings emerging like SEPIA, SECREC, RMIND, FastGC, FairPlay, OblivC etc. A good overview of this could be found here \cite{coed}.

%\subsection{Searchable Encryption}
\section{Threat Intelligence Lookups}
Many free and commercial services offer different types of \emph{IOCs} as watch lists. Primarily there are two modes of operation, either the organizations query these online services on need basis or consume these watch lists as a feed for integrating into their internal analytics.
 
Subsequently they take appropriate action for identification of infections and remediation. Considering the sheer volume of such threat intelligence, many resource constrained organizations would prefer to do a lookup from their client machines to these online services than receiving them as feeds. 

Looking up these Cloud based services needs sharing of \emph{telemetry} which could contain private and sensitive information. Intrusion detection based on threat intelligence is predominantly about identification of \emph{known-known}\footnote{Malware campaigns, Vulnerabilities that are publicly known}  cases in telemetry.
 
\subsection{URL Reputation}
Modern sophisticated attacks like \emph{phishing, domain generated algorithms, command and control communication, data ex-filtration} are predominantly web based \cite{hutchins2011intelligence},\cite{sanskc}. In order to safe guard the users from malicious web based attacks, it is critical to know the reputation of the URL.  Many prominent services exist that provide URL scanning service\cite{urlbl}. Given a particular URL, such services determine whether it is \emph{known bad}. Sharing the web browsing data would severely breach the privacy of individual. In order to prevent such privacy breach,  anonymization techniques proposed for URLs involve encrypting parts of the URL or truncating certain parameters etc. \cite{kuenning2003anonymization}. But such sanitization offers very poor privacy guarantees and are considered weak \cite{pang2006devil},\cite{coull2008taming},\cite{king2009taxonomy}.

Other look up services are malicious public IP addresses\cite{sb}, malware signatures \cite{mlwr} etc.
\subsection{Architecture}
\begin{figure} 
  \begin{center}
    \includegraphics[width=0.5\textwidth]{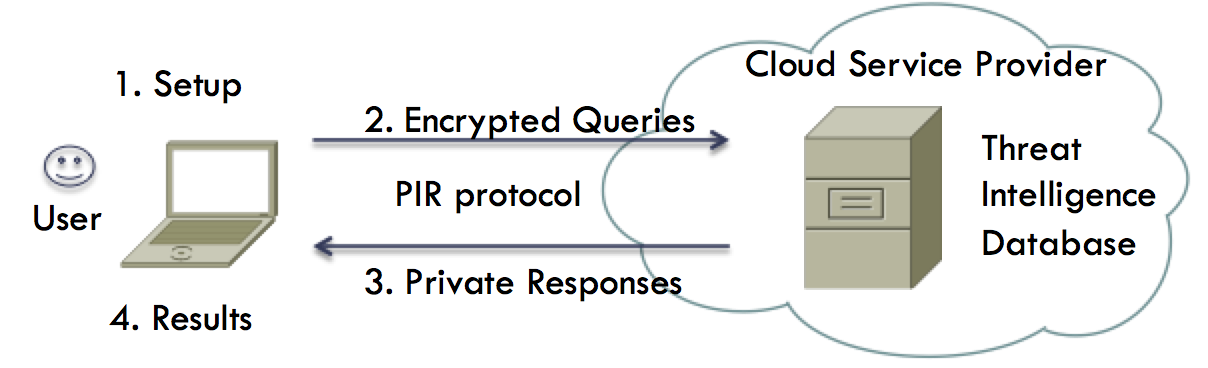}
  \end{center}
  \caption{PIR for Threat Intelligence Lookup}
  \label{fig:pir}
\end{figure}

Privacy preserving threat intelligence lookups is a classic case for \emph{Private Information Retrieval(PIR)}. The users could keep their web browsing history (or network logs) private and lookup a public Cloud service in order to determine if they got infected. The proposed setup is shown in Figure.\ref{fig:pir}
\subsection{Threat Model}
In this setup, privacy of \emph{Users} is ensured from \emph{Cloud Service Provider(CSP)}. The server is oblivious of the lookup strings and records that have been searched. Also \emph{CPIR} protocols needs only one server in contrary to \emph{IT-PIR} protocols that need many non colluding servers. In reality it is hard to prove if the servers of a single \emph{CSP} are non-colluding so \emph{CPIR} protocols are preferred.
\subsection{Challenges}
It has to be experimentally validated to determine how practical current schemes are. The schemes should be efficient many folds than trivial approach of downloading entire threat intelligence database to local network.
%\section{Threat Exchange Platforms}
%\subsection{Threat Model}

\section{Collaborative Sharing and Analytics}
\subsection{Threat Intelligence Sharing}
Many standards are emerging for sharing security related information \cite{mitre}.  Such standards enable organizations to share threat intelligence to prevent attacks in much earlier phases.
Example recommendations that could be made with availability of collective attack observables could be like below

\begin{enumerate}
\item  \emph{40 out of 50 organizations similar to you are hit with this signature}
\item \emph{28 out of 30 organizations in your geography has blocked his \$port or \$ip address or \$domain }%, do you want to ?}
\item \emph {almost 98\% of similar organizations have these top 20 rules as best practices ! " }
\end{enumerate}
Currently projects like DSHIELD helps in identifying trends based on firewall logs.  One of the major concerns still is privacy of information that could be shared in order for these efforts to be effective.  Privacy preserving techniques that enable analysis on such attack observables are still in nascent stages. 

\subsection{Telemetry Sharing}
Telemetry sharing and subsequent security analytics is an essential ingredient of modern intrusion detection for identifying \emph{unknown} cases \cite{bd}. While analytics within the administrative domain only provide local context, global context like ongoing Distributed Denial of Service(DDoS) attacks, passive DNS analysis to identify botnets, Internet wide scans, unusual traffic patterns etc. would require analytics across multiple organizations. 

\subsection{Architecture}
 \begin{figure} 
  \begin{center}
    \includegraphics[width=0.9\textwidth]{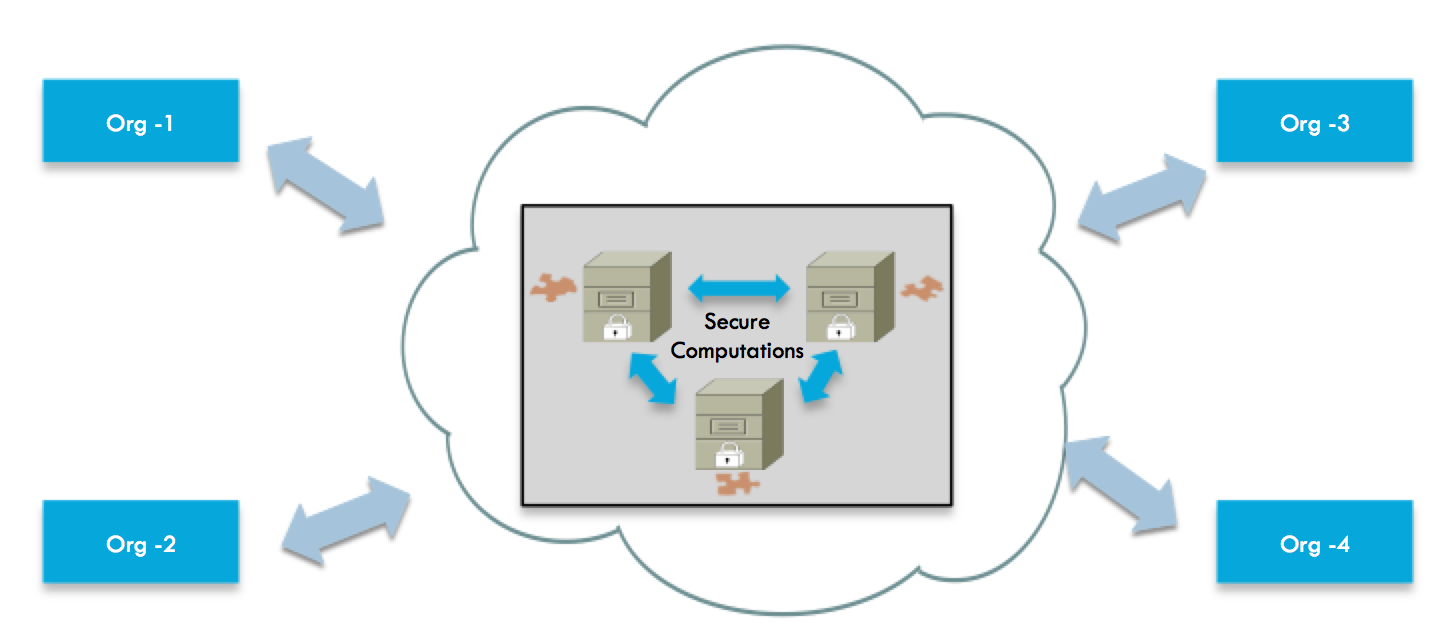}
  \end{center}
  \caption{Collaborative Security Analytics}
  \label{fig:ca}
\end{figure}
This architecture was first proposed for collaborative privacy preserving network security monitoring\cite{burkhart2010sepia}. Network events were aggregated from different administrative domains and protocols were proposed to determine \emph{top-k} nodes in a privacy preserving way. Sensitive network logs and events were \emph{secret shared} and distributed among servers. Subsequently correlation were performed using \emph{Secure Multi party Computation(SMC)} protocols such that no single server has access to entire inputs. Similar approach could be taken for collaborating on \emph{Threat Intel Sharing} and \emph{Telemetry Sharing} as well. 

\subsection{Threat Model}
In this setup none of the participants has access to private inputs of others. The results are made available to all participants as an incentive for participation. The distributed nodes performing \emph{Secure Multiparty Computation (SMC)} can be run be consortium of organizations in a particular field or across fields. For example \emph{Intelligence Sharing and Analysis Centers(ISAC)} of different fields like finance, banking, governments could run the infrastructure needed for \emph{SMC} protocols. Subsequently  the results to the participant organizations could be given as a value add in a privacy preserving way.
\subsection{Challenges}
Recent advances in tools like RMIND\cite{bogdanov2014rmind} are promising. They enable statistical analysis in a privacy preserving manner on \emph{secret shares} of sensitive data to ensure their privacy.  Rigorous experimentation of collaborative algorithms for traffic analysis, botnet detection etc. needs to be performed in order to evaluate their practical viability. 

\section{Conclusions}\label{section:con}
In this position paper we identified privacy preserving architectures for collaborative intrusion detection methods. Various challenges and advances needed in privacy preserving techniques are discussed throughout in realizing such architectures. As future work we intend to experimentally evaluate these proposed architectures in order to validate whether their feasibility in real world. 
% use section* for acknowledgement
\section*{Acknowledgment}
% optional entry into table of contents (if used)
%\addcontentsline{toc}{section}{Acknowledgment}
We would like to  thank Cisco Systems for  supporting this work. 
%We would like to thank {TODO : Insert names }  for many insightful suggestions and improvements in this work. We would also like to acknowledge and thank my immediate management  Anil Nair for supporting me while carrying out this work. 

\bibliographystyle{IEEEtran}

\bibliography{conf}

%\subsection{Challenges}

\end{document}